\documentclass[11pt,twoside]{article}
\usepackage{asp2010}
\usepackage{graphicx}
\usepackage{natbib}
\resetcounters

\begin{document}

\title{The \textit{Suzaku} view of highly-ionised outflows in AGN.}
\author{Jason~Gofford,$^{1}$ James~N.~Reeves,$^{1}$ Valentina~Braito,$^{2}$ Francesco~Tombesi$^{3,4}$
\affil{$^{1}$Astrophysics Group, Keele University, Keele, Staffordshire, ST5 5BG, UK\\
$^{2}$INAF - Osservatorio Astronomico di Brera, Via Bianchi 46, I-23807 Merate, Italy\\
$^{3}$X-ray Astrophysics Laboratory and CRESST, NASA/GSFC, Greenbelt, MD 20771, USA\\
$^{4}$Department of Astronomy, University of Maryland, College Park, MD 20742,
USA\\}}

\begin{abstract}
We are conducting a systematic study of highly-ionised outflows in AGN using archival \textit{Suzaku} data. To date we have analysed 59 observations of 45 AGN using a combined energy-intensity contour plot and Montecarlo method. We find that $\sim36\%$ (16/45) of sources analysed so far show largely unambigous signatures (i.e., Montecarlo proabilities of $>95\%$) of highly-ionised, high-velocity absorption troughs in their X-ray spectra. From XSTAR fitting we find that, overall, the properties of the absorbers are very similar to those found recently by \cite{tombesi10a,tombesi11a} with \textit{XMM-Newton} for the same phenomenon. 
\end{abstract}

\section{Introduction}
Recent systematic studies with \textit{XMM-Newton} have found that as many as $\sim40\%$ of local (i.e., $z<0.1$) AGN show evidence for highly-ionised, high-velocity material outflowing from the central nucleus \citep{tombesi10a}. These outflows -- which are often referred to as ``disk-winds" owing to their origin on sub-parsec scales -- are observationally characterised by blueshifted Fe\,{\sc xxv}~He$\alpha$ and Fe\,{\sc xxvi}~Ly$\alpha$ absorption lines and can have a kinetic energy flux which can exceed the $0.5\%$ of $L_{\rm bol}$ threshold thought necessary to be significant in galaxy scale feedback scenarios \citep{hopkinselvis10}. Furthermore, it is often found that by integrating the kinetic energy flux over a realistic Super-Massive Black Hole (SMBH) e-folding time that the total energy output can be comparable to the binding energy of a typical $10^{11}\,M_{\odot}$ galaxy buldge, which also suggests that they may be energetically significant (e.g., \citealt{reeves10}, \citealt{gofford11}). The study of these outflows is therefore key if we are to bridge the small- and large-scale processes occuring in galactic evolution and, ultimately, to help understand the apparent fundamental relationships observed between galaxy properties and the central nucleus (e.g., the $M_{\rm BH}-\sigma$ relation, \citealt{ferrarese00})  

To this end we are conducting a new systematic study of highly-ionised AGN outflows using archival data for the \textit{Suzaku} space observatory. The wide energy coverage ($0.6-50.0$\,keV) offered by \textit{Suzaku}, in addition to its high throughput and excellent energy resolution at Fe\,K ($\sim120-130$\,eV at $6$\,keV), make it the ideal instrument to simultaneously disentangle the broad-band components of AGN spectra and hence confidently search for any highly-ionised absorption lines. Here, we describe the preliminary results of this work.

\subsection{Sample selection}
The heterogenously selected \textit{Suzaku} sample consists of all pointed AGN observations contained in the Data Archives and Transmission System (DARTS) hosted by the JAXA. Observations were further screened according to the following criteria: (1) Publically available as of September 2011; (2) a net exposure of $>50$\,ks to ensure sufficient photon statistics in even intrinsically faint sources. Observations which were split into multiple sequences (e.g., variability monitoring campaigns) were included if the \textit{total} net exposure was $>50$\,ks; (3) a minimum of $10$\,k counts in the $2-10$\,keV band to detect any highly-ionised absorption lines at a reasonable confidence level. 10k counts is roughly the minimum number required to detect absorption lines at Fe\,K Montecarlo significances of $>95\%$; %This criteria also means those objects where the count rate is dominated in other parts of the spectrum, e.g., the soft-excess or the hard X-ray band, are excluded; 
(4) Compton-thin, i.e., observed $N_{\rm H}<\sigma_{T}^{-1}\approx1.5\times10^{24}$\,cm$^{-2}$, so that the intrinsic continuum emission, and any associated line-of-sight absorption, is directly observable. 
The resultant sample consists of sources which are predominantly local but also includes PDS\,456 ($z=0.184$) and gravitationally lensed quasar APM\,08279+5255 ($z=3.9122$).

\section{Analysis}
\subsection{Fitting the continuum}
The X-ray spectrum of AGN typically consists of a primary power-law continuum, Compton reflection, and, in some cases, a soft excess. When absorbed by distant photoionised material these components can impart significant complexity into the Fe\,K band (e.g., the Fe\,K$\alpha$ fluorescence line and associated edge). It is therefore important to first constrain the broadband continuum, and the effects of each of the spectral components to the Fe\,K band, before assessing for the presence of Fe\,{\sc xxv}~He$\alpha$ and/or Fe\,{\sc xxvi}~Ly$\alpha$ absorption.

We utilise the full bandpass of the \textit{Suzaku} XIS+PIN (i.e., $0.6-50$\,keV) to constrain the contribution of the various continuum components, and, using table models generated using the XSTAR photoionisation code \citep{bautista_xstar}, model the considerable effects that both fully- and partially-covering absorption can have on the broad-band spectral curvature and the resultant continuum parameters. In particular, we give close attention to the Fe\,K edge at $\sim7.1$\,keV and account for the effects that both Compton reflection \textit{and} photionised absorption can have on the edge profile. The Fe\,K$\alpha$ fluorescence line and associated reflection hump are characterised with REFLIONX which calculates the reflected emission from a face-on slab of Compton-thick material \citep{ross_fabian_reflionx}. %Importantly, this allows us to break any possible degenercy between power-law slope, $\Gamma$, and reflection fraction, $R$, and ensures that contribution of reflection between $5-10$\,keV is taken into account before searching for any highly-ionised absorption lines. 
In the soft X-ray band XSTAR absorption tables are added until an adequate fit to the spectrum is achieved.

\subsection{Searching for atomic features} 
For a given best-fit model, i.e., accounting for all continuum components, soft X-ray emission lines and any warm absorption, we have searched for ionised atomic features in the Fe\,K band using unbiased and statistically driven energy-intensity contour plots similar to those used by \citep{tombesi10a}. The contour plots are generated by stepping a Gaussian in $25$\,eV intervals between $5.0-10.0$\,keV (rest-frame), with the line normalisation allowed to adopt both positive and negative values. The resultant plots provide a powerful means to visually determine whether there are any possible atomic features present between $5-10$\,keV which warrent further invesitgation with Montecarlo simulations (see section \ref{sec:montecarlo}).  Energy-intensity contour plots were generated for each observation in the sample and all residuals with an F-test significance $>99\%$ were then modelled with Gaussian profiles. If a spectrum appears to require a broadened Fe\,K$\alpha$ component the residua were modelled with a broad Gaussian or a DISKLINE depending on which yielded the best phenomenological fit to the data. In these cases a second contour plot was generated -- this time with the broad component included -- to determine whether its presence had any effect on the absorption line detection. 

All residua which were found to have F-Test and Montecarlo significances of $P_{\rm F}>99\%$ and $P_{\rm MC}>95\%$, respectively, were self-consistently modelled using an XSTAR photoionisation table with a turbulent velocity, $v_{\rm turb}$, roughly matching the full-width at half-maximum (FWHM) velocity width of the best-fit Gaussian line. The use of XSTAR enables us to measure the key parameters of the absorbing material, such as the ionisation parameter, $\xi$, column density, $N_{\rm H}$, and outflow velocity, $v_{\rm out}$, relative to the observer. Importantly, this then allow us to indirectly probe how the likely energy budget of the outflow and assess whether it is likely to play a significant role in galaxy-wide feedback scenarios.

\begin{figure}[!t]
\plottwo{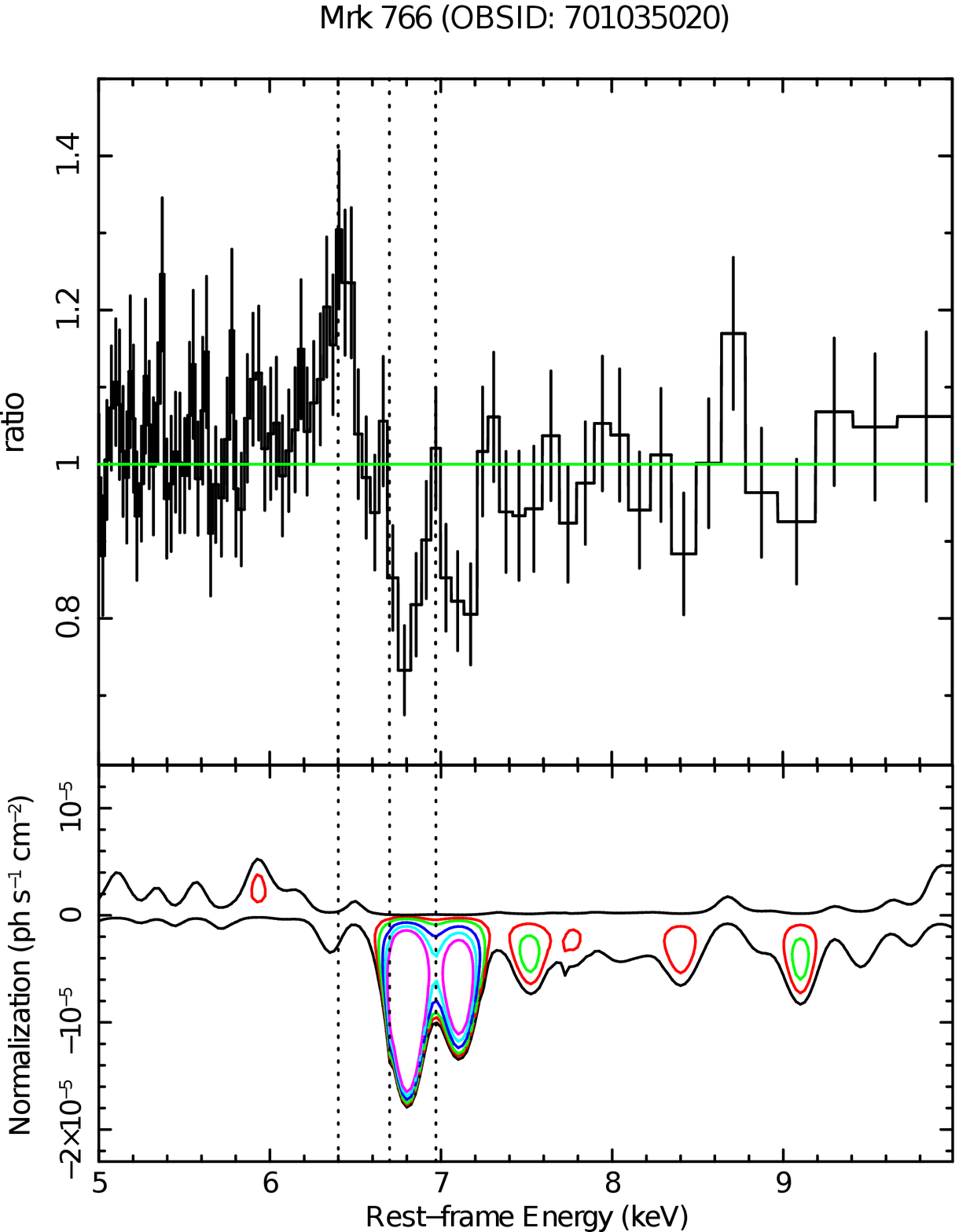}{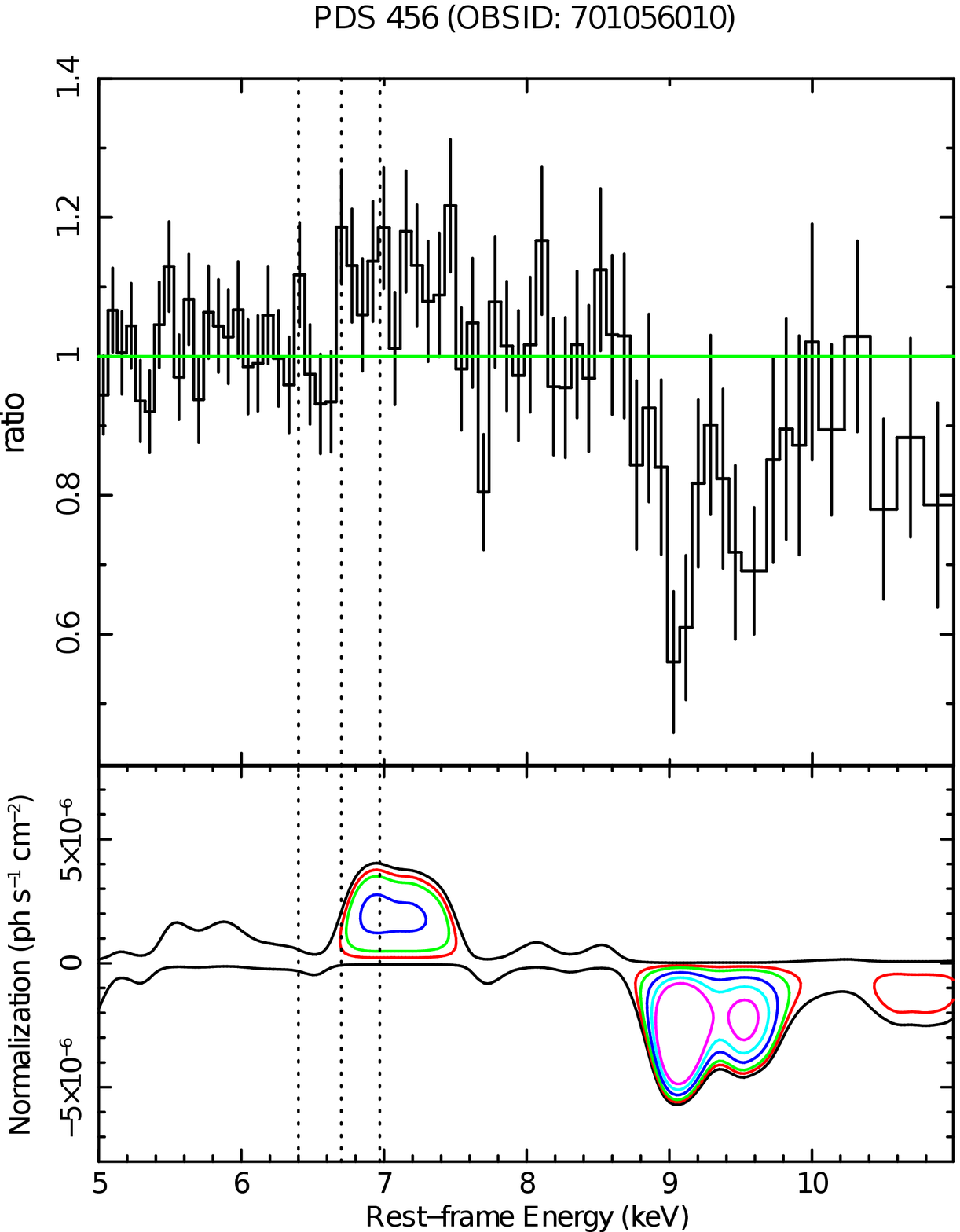}
\caption{Data/model residuals and energy-intensity contour plots for Mrk\,766 (left) and PDS\,456 (right). The data/model residuals are given to the best-fit continuum model of each source with reflection modelled using the PEXRAV model to highlight the Fe\,K$\alpha$ emission line. The continuous open contour corresponds to a $\Delta\chi^{2}=+0.5$ and is intended to indicate the level of the continuum. From outer to inner, the closed contours %show $\Delta\chi^{2}$ improvements of -2.3, -4.61, -9.21, -13.82, -18.42. These 
correspond to F-test significances of $68\%$, $90\%$, $99\%$, $99.9\%$ and $99.99\%$, respectively. The dashed vertical lines indicate the expected rest-frame energies of the Fe\,K$\alpha$, Fe\,{\sc xxv}~He$\alpha$ and Fe\,{\sc xxvi}~Ly$\alpha$ lines.}
\label{fig:ratcont}
\end{figure}

In Figure \ref{fig:ratcont} we show some examples of this fitting process when applied to two objects in the sample: Mrk\,766 and PDS\,456. The energy-intensity contours show that there is unambiguous evidence for blueshifted absorption at F-test confidences of $>99.99\%$ in both sources. In Mrk\,766 (left panel) the two profiles are unresolved ($\sigma=10$\,eV) and have an energy separation consistent with He$\alpha$ and Ly$\alpha$. Both lines are fit with a single highly-ionised absorption zone described by $N_{\rm H}=(5.6^{+1.5}_{-1.4})\times10^{22}$\,cm$^{-2}$ and $\log\xi=3.67\pm0.06$, with a modest blueshifted velocity of $v_{\rm out}=-(5.4\pm0.9)\times10^{3}$\,km\,s$^{-1}$ relative to the systematic. Conversely, the absorption lines in PDS\,456 (right) are much more extreme. Our best-fit XSTAR model has two absorbers with tied $N_{\rm H}=(1.50^{+0.25}_{-0.30})\times10^{23}$\,cm$^{-2}$ and $\log\xi=4.26^{+0.13}_{-0.14}$, and outflow velocities of $v_{\rm out,1}=-(81.0\pm2.1)\times10^{3}$\,km\,s$^{-1}$ and $v_{\rm out,2}=-(90.0\pm2.4)\times10^{3}$\,km\,s$^{-1}$, respectively. Interestingly, this suggests that both lines are due to blends of He$\alpha$/Ly$\alpha$ and, therefore, that the the disk-wind may be clumpy and have multiple components with different outflow velocities. PDS~456 is one of the most extreme objects in the sample.

\subsection{Montecarlo simulations}
\label{sec:montecarlo}
\cite{protassov02} showed that the F-test yields inaccurate statistics when dealing with complex spectral models containing narrow spectral lines. We, therefore, do not claim to robustly detect any lines based solely on the F-Test. Instead, we take the F-test probability, $P_{\rm F}$, as a rough `guide' when analysing the spectum and follow up all suspected absorption lines that have $P_{\rm F}>99\%$ with a Montecarlo (MC) analysis designed to determine their true veracity against random noise in the spectrum. 

The MC method is as follows: from the null model, i.e., no ionised lines present, we simulate $T=1000$ fake XIS (FI) spectra using the FAKEIT command in XSPEC. In \textit{each} of these simulated spectra a Gaussian profile with width equal to that of the suspected line is stepped every 25\,eV between $5-9.5$\,keV (180 steps) with the centroid energy free to vary and the line normalisation allowed to adopt positive and negative values. A global $\Delta\chi^{2}$ distribution was then generated by recording the \textit{maximum}  improvement due to random noise, $\Delta\chi^{2}_{\rm noise}$, when stepping the line through each simulated pectrum. If a Gaussian profile in the real spectrum gives a $\Delta\chi^{2}_{\rm real}$ improvement, the MC probability of this line is then $P_{\rm MC}=1-(N/T)$, where $N$ is the number of simulated spectra with $\Delta\chi^{2}_{\rm noise}>\Delta\chi^{2}_{\rm real}$, and $T$ is the total simulated spectra.

\section{Results}

\begin{figure}[t]
\plotthree{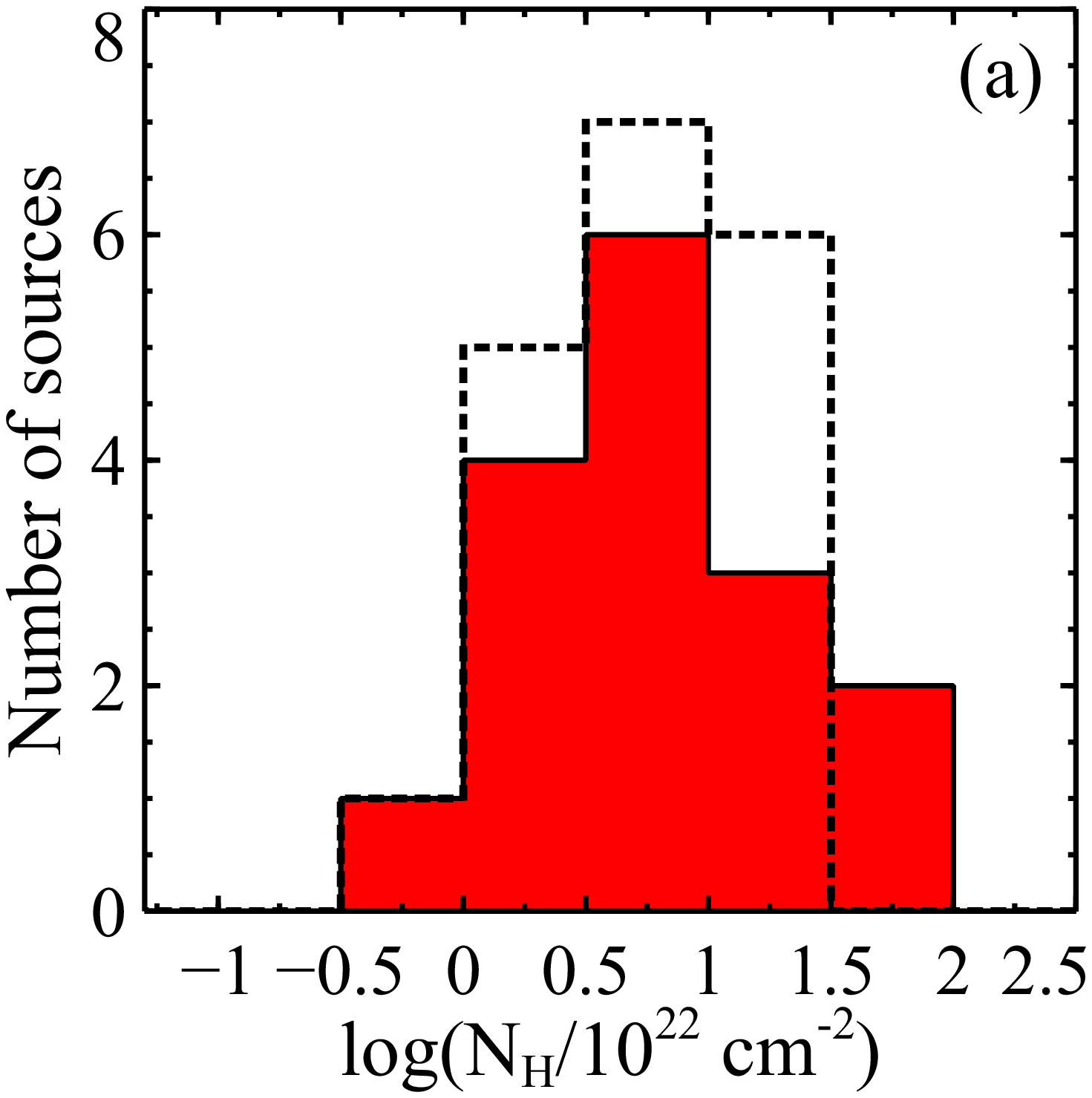}{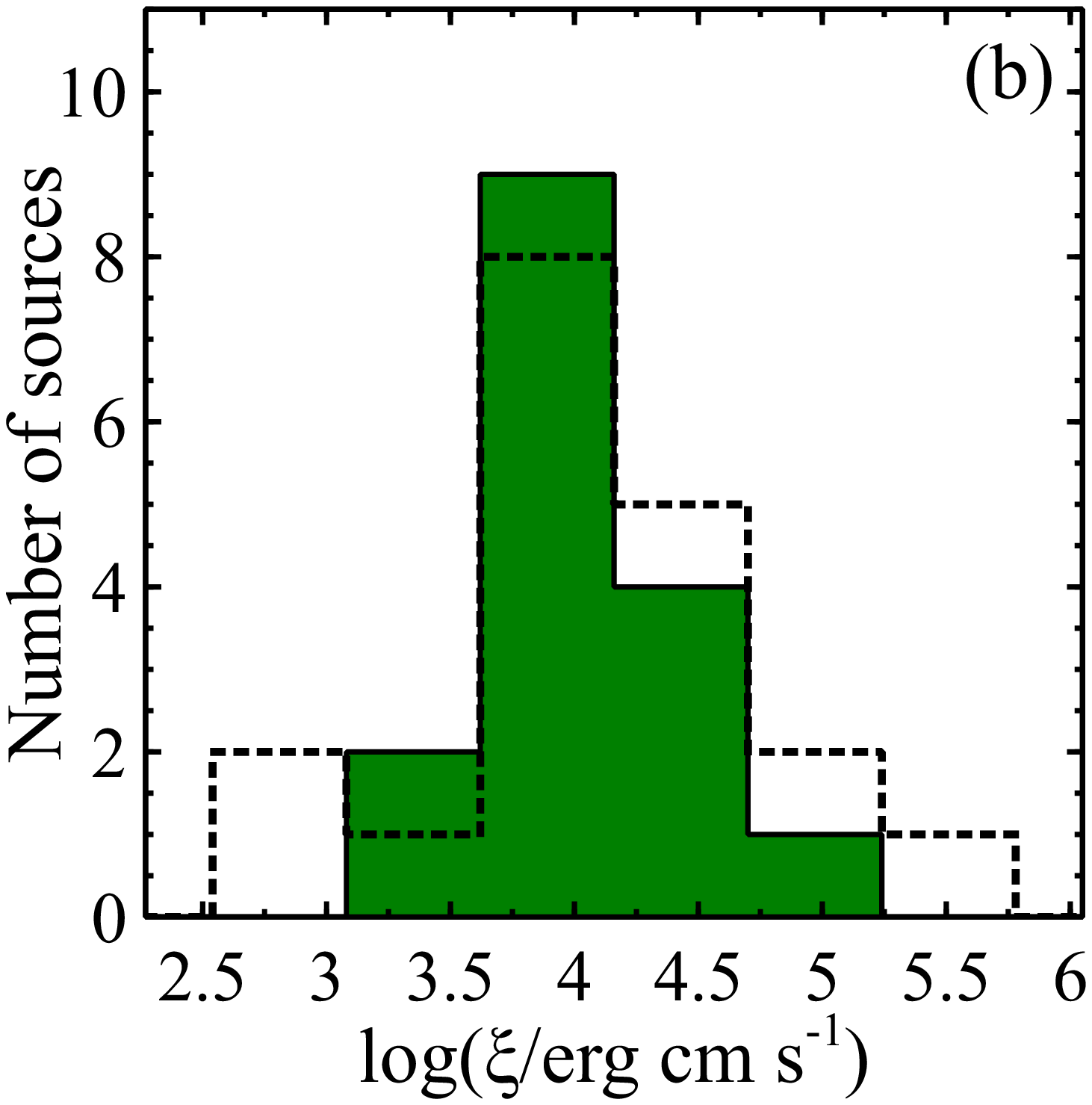}{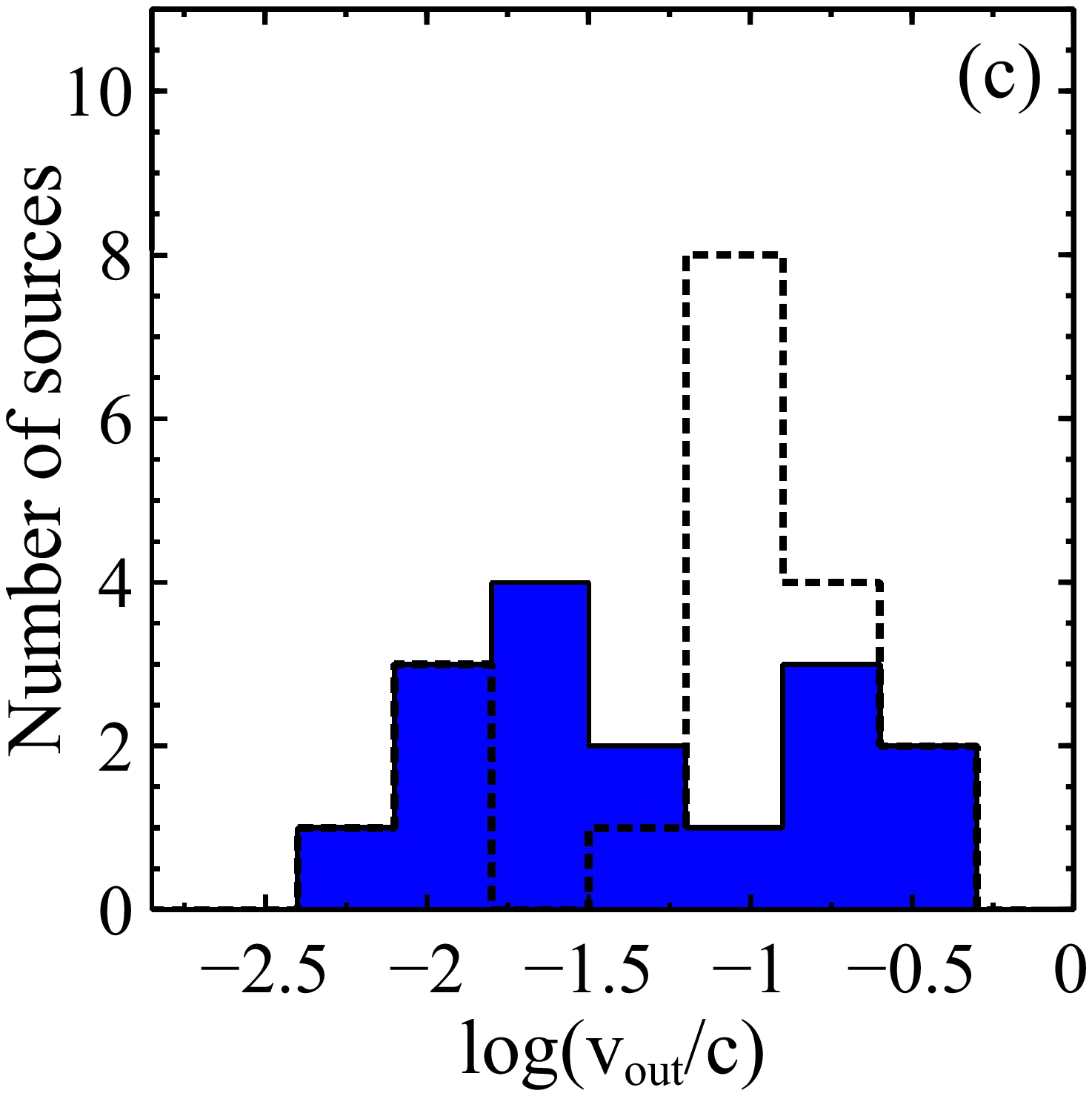}
\caption{Histograms showing the distribution of absorber parameters: (a) column density in units of $\log(N_{\rm H}/10^{22})$; (b) ionisation parameter in units of $\log(\xi/\rm{erg\,cm\,s}^{-1})$; (c) outflow velocity in units of $\log(v_{\rm out}/c)$. In all panels the shaded/coloured area corresponds to this work while the dashed line shows the results of \cite{tombesi10a}.}
\label{fig:xstarresults}
\end{figure}

%\subsection*{Absorber properties}
While work on the sample is ongoing we have fully analysed 45 AGN with this method. In these objects 16/45 (21/59 observations) show evidence for statistically significant ($P_{\rm MC}>95\%$) highly-ionised absorption lines. This corresponds to $\sim36\%$ overall detection rate and is very similar to the global detection fraction found by \cite{tombesi10a} in local radio-quiet Seyfert galaxies. 

The observational signatures for the outflows falls into four categories: the most frequently observed absorbers are those consisting of either a single Fe\,{\sc xxvi} Ly$\alpha$ line or both a Fe\,{\sc xxv}\,He$\alpha$/Fe\,{\sc xxvi} Ly$\alpha$ pair, which each accounting for 6/16 ($\sim38\%$) of the detected outflows. A further 2/16 ($\sim 13\%$) are most likely due to Fe\,{\sc xxv} He$\alpha$ blended with lower ionisation species of iron, and the final 2/16 have more than one absorption profile (i.e., PDS\,456 in fig. \ref{fig:ratcont}) which suggests the presence to a multi-velocity system. The current distributions of XSTAR absorber parameters are shown in fig. \ref{fig:xstarresults}. Overall, the results are in good agreement with those of the \cite{tombesi10a,tombesi11a} \textit{XMM-Newton} sample. The $N_{\rm H}$ values [panel (a)] in particular are very similar and, while the measured column densities cover a wide range (i.e., $10^{21} \la N_{\rm H}/$\,cm$^{-2} \la 10^{24}$) there is a peak at $(3-10)\times10^{22}$\,cm$^{-2}$ and a mean value of $N_{\rm H,suzaku}\approx1\times10^{23}$\,cm$^{-2}$. This is consistent with the analagous values found with \textit{XMM-Newton}. The distribution of ionisation parameter [panel (b)] covers the range range $3.1 \la \log(\xi/\rm{erg\,cm\,s}^{-1}) \la 5.5$ and, with a peak and mean value at $\log(\xi/\rm{erg\,cm\,s}^{-1}) \approx 4.0$, is again consistent with the results found by \cite{tombesi10a}. Despite this good agreement in terms of intrinsic absorber properties the current distrubtion of outflow velocities appears to differ somewhat, as is shown in panel (c). Whereas there is a sharp peak at $\sim0.1$\,c found with \textit{XMM-Newton}, we find a smoother and more continuous range of velocities with \textit{Suzaku}, ranging from $0.004 \la v_{\rm out}/\rm{c} \la 0.5$, with very few at the expected peak velocity of $0.1$\,c (e.g., see \citealt{king10}). However, given that analysis of the sample is ongoing it is currently too soon to determine whether this is an intrinsic property of the absorbers or a result of, for example, low number statistics or intrinsic detector biases. The full \textit{Suzaku} outflow sample and a thorough discussion regarding all of the work outlined here will be given in Gofford et al. (2012; in preparation).

\bibliography{gofford}

\end{document}